# High pressure synthesis of a new superconductor $Sr_2CuO_{2+\delta}Cl_{2-y}$ induced by "apical oxygen doping"

Q.Q. Liu, X. M. Qin, Y. Yu, F. Y. Li, C. Dong, C. Q. Jin*

Institute of Physics, Chinese Academy of Sciences, P. O. Box 603, Beijing 100080, P. R. China

## Abstract

Using the apical oxygen doping mechanism, i.e. a partial substitution of divalence O for the monovalence Cl, a p-type oxychloride cuprate superconductor, $Sr_2CuO_{2+\delta}Cl_{2-y}$, was synthesized at high pressure high temperature. The x-ray diffraction refinement suggests the superconductor crystallizes into a 0201 structure with space group I4/mmm and lattice parameters being $a$=3.92Å, $c$=15.6 Å. The magnetic susceptibility as well as resistance measurements indicated that the bulk superconductivity with transition temperature ($T_c$) 30K was achieved in the sample.

* Corresponding author

# Introduction

The crystal structure of high Tc superconductors (HTS) is built of two blocks, i.e., the charge reservoir and [CuO2] conducting layer, while the connection between the charge reservoir and the [CuO2] conducting layer is the apical oxygen. There have been intensive interests in studying the apical oxygen of cuprate superconductors[1,2]. Isostructural to the oxide counterpart, various oxychloride cuprate compounds with apical chlorine including $(Sr,Ca)_2CuO_2Cl_2$ with the 0201 ($K_2NiF_4$ type) structure, and $Ca_3Cu_2O_2Cl_2$ with the 0212 (or $La_2SrCu_2O_6$ type) structure are ideal systems to investigate the role of apical anion atoms in HTS. Since the divalence oxygen is heterovalence with the monovalence chlorine, it is possible to induce hole carriers into the [CuO2] plane of the compounds simply through the substitution $O^{2-}$ for $Cl^-$, namely by "apical oxygen doping" to reach high Tc superconductivity [3]. This was first realized in the double [CuO2] layered copper-oxychloride compound $(Sr,Ca)_3Cu_2O_{4+\delta}Cl_{2-y}$. The [CuO2] plane in oxychloride cuprate is geometrically more two dimensional due to the larger c-axis distance of the chare reservoir block. It was theoretically predicted that the superconductivity in these materials could be induced when doped by sufficient holes per unit cell [4]. Oxychloride cuprates have received considerable interests by physicists as ideal two dimensional model compounds of HTS. For instance, recently series experiments of angle resolved photoemission spectroscopy (ARPES) have been conducted on the 0201 type oxychloride cuprates [5-7]. More recently an antiferromagnetism and hole pair checkerboard model was theoretically proposed to understand the possible evolution from a Mott insulator to a superconducting regime with increasing carrier doping in HTS [8]. This electronic crystal state was experimentally first observed in the 0201 type oxychloride cuprate $(Ca,Na)_2CuO_2Cl_2$[9] Therefore investigating these type oxychloride cuprates may provide an important clue to learn high Tc superconducting mechanism.

High-pressure synthesis has long been considered an effective tool in searching novel high $T_c$ superconductors, especially stabilizing the superconducting homologous series [10-15]. It is noted that so far all copper oxychloride superconductors [3,15-20] have to be synthesized using a high-pressure and high-temperature technique. For example, $(Sr,Ca)_3Cu_2O_{4+}Cl_{2-y}$ superconductor with $T_c$ 80K of 0212-type structure was synthesized by using the "apical oxygen doping", under high pressure[3]. Subsequently, the 0223-type Sr-Ca-Cu-O-Cl superconductor with $T_c$ 35K, which is the higher member of the same homologous series Cl-02(n-1)n, has also been synthesized using the same doping mechanism under high pressure[19]. The synthesis of these materials is helpful to investigate the role of apical oxygen atoms in high-$T_c$ cuprate superconductors and to identify essential features between the charge reservoir block and the [CuO2] conducting block common to all HTS.

The parent oxychloride $Sr_2CuO_2Cl_2$ with 0201-type structure was previously prepared under ambient pressure [21]. The crystal structure model is illustrated in Fig.1a, and that of $La_2CuO_4$ is drawn in Fig.1b for comparison. In 1996, Hiroi et al. [22] attempted to dope $Sr_2CuO_2Cl_2$ with sodium and potassium, but there was no indication of substitution or hole carrier density. Later, T.Tatsuki, et al. [23] synthesized 0201 type $Sr_2CuO_{2+}Cl_{2-y}$ compounds with oxygen partially substituting for the apical chlorine sites by using high-pressure technique, and confirmed the apical sites partially occupied by oxygen, which introduced holes into the [CuO2] sheets. However, the superconductivity was absent, and they speculated that it was caused by the insufficient amount of mobile holes in the [CuO2] sheets of $Sr_2CuO_{2+}Cl_{2-y}$.

Here, we report that for the first time $Sr_2CuO_{2+}Cl_{2-y}$ superconductors were synthesized by using the apical oxygen doping mechanism under high temperature and high pressure. Their structure and superconductivity were measured. Bulk superconductivity with a maximum Tc value of 30 K was observed.

**Experimental**

The samples were synthesized in two steps. At the first step, precursors of $Sr_2CuO_3$, $Sr_2CuO_2Cl_2$ were prepared by conventional solid state reaction method from the mixture of $SrCO_3$, $SrCl_2$ and CuO powders with purities higher than 99.9% for each raw material. The powder mixture in an appropriate ratio was ground in an agate mortar and then heated in the range of 900 to 950°C in air for 24h. The processes were repeated several times in order to get the single phase product. At the second step, the precursors were mixed with $SrO_2$ (99.9% pure) and CuO with the nominal composition of $Sr_2CuO_{2+}Cl_{2-y}$ in a dry glove box. The materials were then subjected to high pressure synthesis under 6 GPa pressure and at 1050°C for 30min using a cubic-anvil-type high pressure facility. Here the $SrO_2$ in the starting mixture was also used as an oxidizer to generate an oxygen atmosphere in the high-pressure synthesis, which is needed in order to induce "apical oxygen" during the synthesis.

The structure of the samples was analyzed by means of X-ray powder diffraction (XRD) using Cu K$\alpha$ radiation. The lattice parameters of $a$ and $c$ axis were determined using a Rietveld analysis software (Winplotr). DC electrical resistivity was measured by the standard four-probe method. The DC magnetic susceptibility was determined using a SQUID magnetometer at a 20 Oe applied field in both zero-field cooling (ZFC) and field-cooling (FC) modes.

Results and discussion

By controlling oxygen pressure provided by the oxidizer amount and changing chlorine content, we prepared the superconducting samples with the formula of $Sr_2CuO_{2+}Cl_{2-y}$ for 0.6 y 1.2, Fig.2 shows the XRD pattern for $Sr_2CuO_2Cl_2$ prepared at ambient pressure and for $Sr_2CuO_{2+}Cl_y$ synthesized under high pressure with y=0.8, 1.0, 1.2, 1.4 and 1.6. It was found that most of the peaks in the superconducting samples can be indexed into the $K_2NiF_4$ structure, and small amounts of impurity phases of $SrO_2$ and unknown minor phase(s) were also included in the samples. The crystal structure was determined to be tetragonal, essentially the same as that of the parent material $Sr_2CuO_2Cl_2$. Assuming the tetragonal structure, we refined the cell parameters of the sample and yielded the lattice parameters. Table 1 lists the results of the Rietveld refinements for the $Sr_2CuO_{2+}Cl_{1.2}$ superconductor. The lattice parameters, $a$ and $c$, are plotted as a function of nominal Cl content for the $Sr_2CuO_{2+}Cl_y$ (y = 0.8, 1.0, 1.2, 1.4, 2.0) samples in Fig. 3. As compared with the ambient pressure phase $Sr_2CuO_2Cl_2$ ($a$=3.9716(2) Å, $c$=15.6126(2) Å)[21], the observed shrinkage in the $a$-axis should be resulted from the hole doping into the [CuO2] plane. The increase of c-axis length can be explained by the expansion of the Sr-Cl rock-salt block due to enhanced Coulomb repulsion between the adjacent chlorine layers resulted from the introduction of divalent oxygen.

The superconducting properties of the sample were measured by DC susceptibility using a SQUID magnetometer and electrical resistance by the standard four-probe resistance method. Fig.4 show the temperature dependence of the DC magnetic susceptibility and the electrical resistivity of $Sr_2CuO_{2+}Cl_{1.2}$, respectively. A superconducting transition was clearly observed with Tc~30K from both magnetic and electrical conductivity measurements. The DC susceptibility measurements were carried out in both modes of zero-field cooling and field cooling. The FC data which corresponds to the Meissner signal revealed a superconducting volume fraction ~10% at about 10K.

Fig. 5 shows the temperature dependence of the susceptibility for the samples with y = 0.8, 1.0, 1.4 in FC mode. The superconducting transitions are observed in those samples with $T_c$ 33, 20, 30K for y = 0.8, 1.0, 1.4, respectively. Fig. 6 shows the temperature dependence of the electrical resistivity of the samples. The higher $T_c$ of ~ 36K with y=1.4 may imply that the sample is chemically inhomogeneous. Due to the difficulty in accurately monitoring the synthesizing condition, mapping the change of $T_c$ with apical oxygen doping level has not be qualitatively analyzed yet. Further optimization of the synthesis condition is necessary to obtain more systematic complementary data.

It is worthwhile to note that the $a$-axis parameter of the $Sr_2CuO_{2+}Cl_y$ superconductor is substantially larger than those for many hole-doped cuprate superconductors for which the $a$ is usually less than 3.90 Å and averagely being ~3.83Å[24, 25, 26]. To our knowledge this was so far the first case that a p-type cuprate superconductor shows the Cu-O-Cu distance in the [CuO2] plane larger than 3.90 Å. It is once more an indication of the uniqueness of oxyhalide superconductors in comparing with the pure copper oxide HTS. Some compounds with the same 0201-type or similar structure (1201-type) exhibited insulating properties [27-30], which have larger *a*-axis than 3.90Å and contained large oxygen vacancy (20-75%) in the [CuO2] plane, resulting in the imperfect [CuO2] layers. In contrast, the Rietveld refinements shown in Table 1 suggests that the 0201-type $Sr_2CuO_{2+\delta}Cl_{2-y}$ contained few oxygen vacancies, indicating that the [CuO2] sheet is rather perfect. So in order to induce superconductivity in the $Sr_2CuO_{2+\delta}Cl_{2-y}$ samples, the key point is how to dope appropriate amount of mobile carriers into the [CuO2] sheets which should be perfect and keep the chemical stoichiometry. In addition, the energy dispersive spectroscopy analysis revealed that the average atomic ratio of Sr:Cu:Cl over tens of well grown grains was basically in agreement with the suggested $Sr_2CuO_{2+\delta}Cl_y$ formula. At the same time, high resolution transmission electron microscopy and electron diffraction investigation on a large amount of crystals of the samples did not show the trace of other phases with the similar $K_2NiF_4$ structure [31]. So it is inferred that the observed ~ 30 K bulk superconductivity is caused by the apical oxygen doped Cl-0201 phase $Sr_2CuO_{2+\delta}Cl_{2-y}$.

## Summary

In conclusion, the Cl-0201 type $Sr_2CuO_{2+\delta}Cl_{2-y}$ superconductors were successfully synthesized at 1050°C under 6.0GPa by using "apical oxygen doping" mechanism which introduced mobile holes to the [CuO2] plane. The X-ray powder diffraction showed that the crystal had the hole doped tetragonal 0201-type structure with the lattice parameter *a* larger than 3.90 Å. Bulk superconductivity of Tc 30 K was observed from the DC susceptibility and resistivity measurements.

## Acknowledgement

This work was partially supported by the national nature science foundation of

China (grant numbers 50401009, 50328102, 50321101), the state key fundamental research project (2002CB613301), and Chinese Academy of Sciences.

Table 1: Results of structure refinements for the superconductor $Sr_2CuO_{2+\delta}Cl_{1.2}$ using the Rietveld method.

Space group: I4/mmm; Unit-cell dimensions: $a=b=3.9435(2)$Å, $c=15.6426(1)$Å; $R_{wp}=14.5\%$, $R_p=15.2\%$, $R_{exp}=3.7\%$; The refinement range of $2\theta$ is 5-120º; Cu $K_{\alpha 1}$ radiation was used.

| Atom | Site | X | Y | Z | Occupancy |
|------|------|---|-----|----------|-----------|
| Cu   | 2a   | 0 | 0   | 0        | 1         |
| Sr   | 4e   | 0 | 0   | 0.3940(7)| 1         |
| O1   | 4c   | 0 | 0.5 | 0        | 1         |
| O2   | 4e   | 0 | 0   | 0.141(2) | 0.37(2)   |
| Cl   | 4e   | 0 | 0   | 0.138(6) | 0.63(3)   |

**Figure captions:**

Fig.1 Schematic view of the crystal structure of (a) $Sr_2CuO_2Cl_2$ and (b) $La_2CuO_4$

Fig.2 X-ray powder diffraction pattern for the parent sample $Sr_2CuO_2Cl_2$ and the apical oxygen doped $Sr_2CuO_{2+\delta}Cl_y$. The major phase can be indexed into the Cl-0201 structure. The impurity phases were identified to be $SrO_2$ (•), and unknown phase(*).

Fig. 3. Lattice parameters *a* (open square*)* and *c* (solid circle) versus Cl content (x) of $Sr_2CuO_{2+\delta}Cl_y$ with variant apical oxygen doping.

Fig. 4. Temperature dependence of the DC magnetic susceptibility of the sample $Sr_2CuO_{2+\delta}Cl_{1.2}$ in an applied external field of 20Oe in both ZFC and FC modes. The superconducting transition temperature is 30K as indicated. The calculated superconducting volume fraction according to the Meissner signal is ~10%, indicating the bulk superconducting nature. Inset is the temperature dependence of electrical resistivity measured by the four-probe method.

Fig. 5. Temperature dependence of the magnetic susceptibility at FC mode for the samples of $Sr_2CuO_{2+\delta}Cl_y$ superconductors with variant apical oxygen doping.

Fig. 6. Temperature dependence of the electrical resistivity for the samples of $Sr_2CuO_{2+\delta}Cl_y$ superconductors with variant apical oxygen doping.

Apical chlorine    Apical oxygen

[CuO2] plane

Sr
Cu
Cl
O

La
Cu
O

a

b

**Fig.1/6**

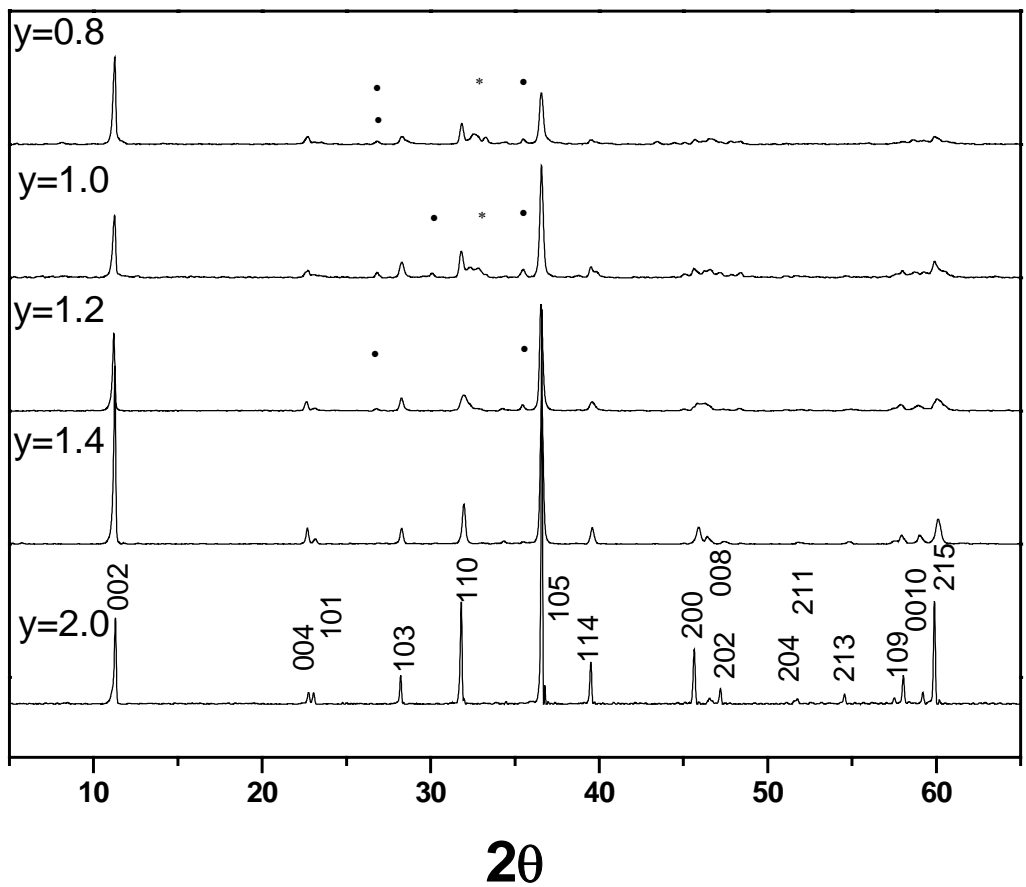

**Fig.2/6**

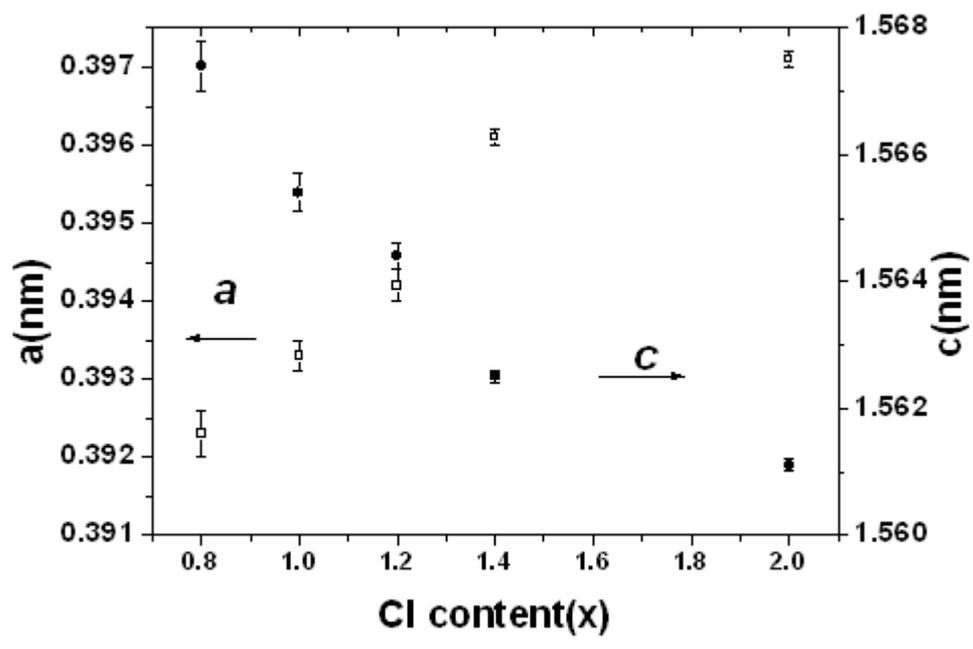

Fig.3/6

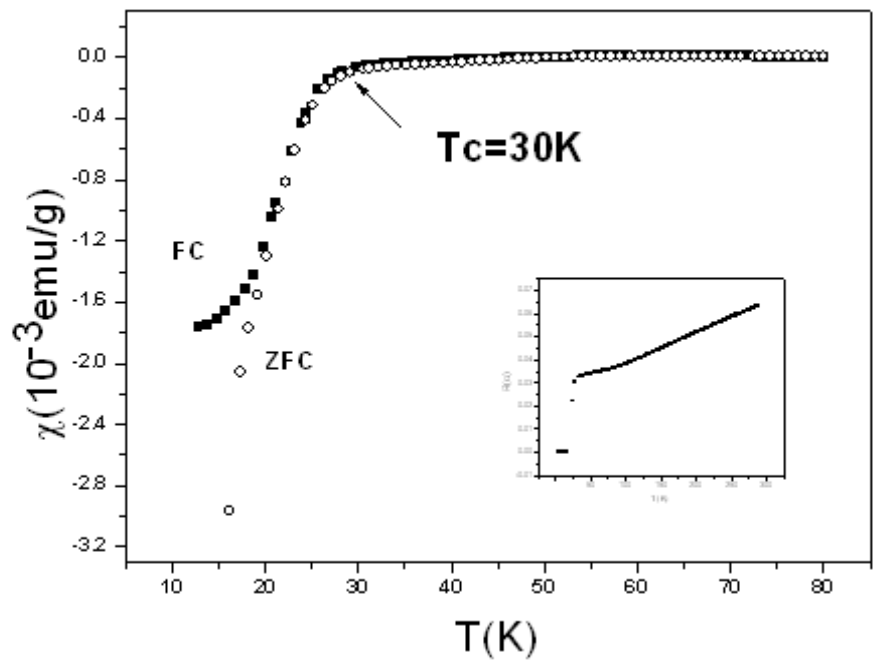

Fig.4/6

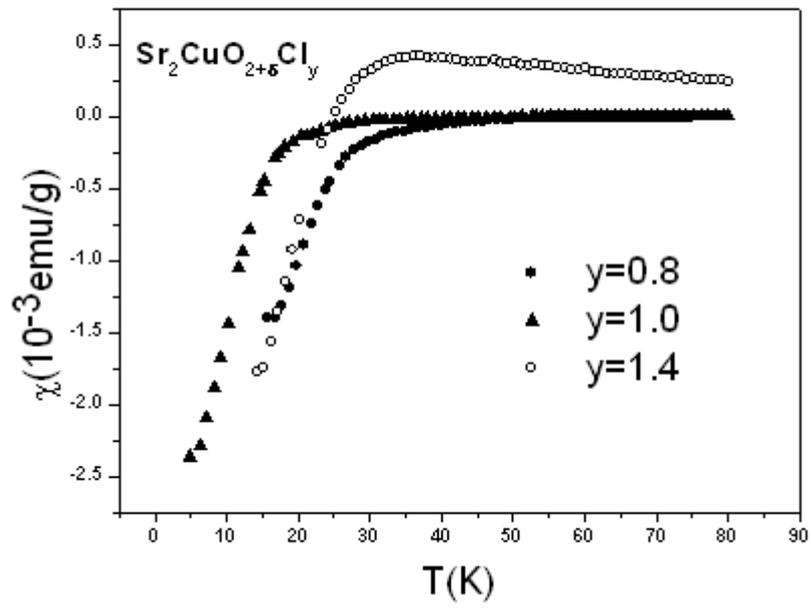

**Fig.5/6**

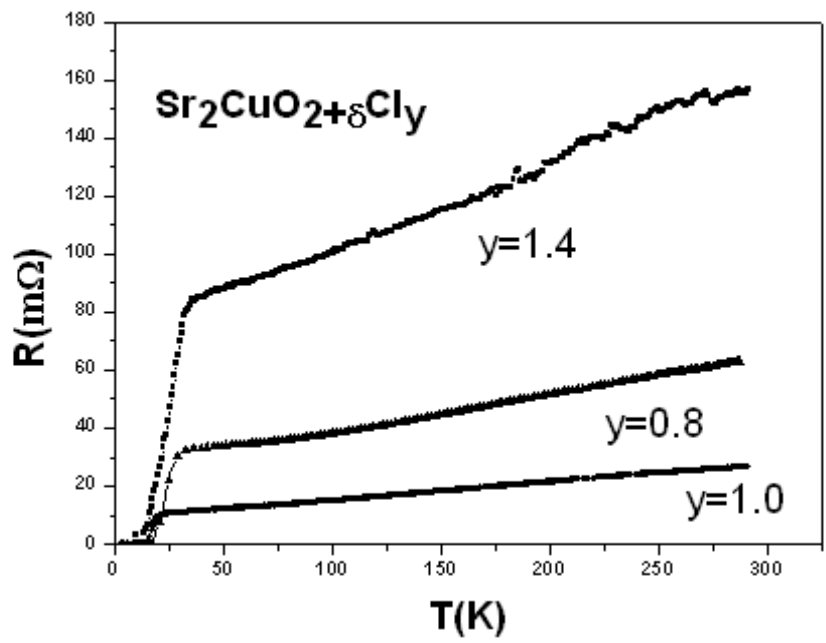

Fig.6/6